\documentclass[aps,prl,twocolumn]{revtex4}
\usepackage{amssymb}
\usepackage{graphicx}
\usepackage{hyperref}

\begin{document}

\title{Dynamics of Magnetic Defects in Heavy Fermion LiV$_2$O$_4$ \protect\\ 
from Stretched Exponential $^7$Li NMR Relaxation}
\author{D.~C. Johnston}
\author{S.-H.~Baek}
\altaffiliation[Present address: ]{National High Magnetic Field Laboratory,
Tallahassee, Florida 32310}
\author{X.~Zong}
\author{F.~Borsa}
\author{J.~Schmalian}
\author{S.~Kondo}
\altaffiliation[Present address: ]{Materials Laboratories, Sony Corporation,
Haneda TEC., 5-21-15 Higashikojiya, Ota-ku, Tokyo 144-0033, Japan.}
\affiliation{Ames Laboratory and Department of Physics and Astronomy, Iowa State
University, Ames, Iowa 50011}
\date{March 21, 2005}

\begin{abstract}
$^7$Li NMR measurements on $\mathrm{LiV_2O_4}$ from 0.5 to 4.2 K are reported. A
small concentration of magnetic defects within the structure drastically changes
the $^7$Li nuclear magnetization relaxation versus time from a pure exponential as
in pure LiV$_2$O$_4$ to a stretched exponential, indicating glassy
behavior of the magnetic defects.  The stretched
exponential function is described as arising from a distribution of $^7$Li nuclear
spin-lattice relaxation rates and we present a model for the distribution in
terms of the dynamics of the magnetic defects.  Our results explain the origin of
recent puzzling $^7$Li NMR literature data on $\mathrm{LiV_2O_4}$ and our model is
likely applicable to other glassy systems.
\end{abstract}

\maketitle

Heavy fermion (HF, heavy Fermi liquid) behaviors have been widely observed
at low temperatures $T$ in many metals containing crystallographically
ordered arrays of $f$-electron atoms, which are quite well
understood theoretically \cite{Hewson1993}.  In these metals, the current
carriers  act as if they have a (heavy) mass that is of order 10$^{2}$--10$^{3}$
times the free electron mass.  Only a few $d$-electron compounds are known to show
HF behaviors at low $T$, e.g.\ Y$_{1-x}$Sc$_{x}$Mn$_{2}$ with $x\approx
0.03$ \cite{Ballou1996}, $\mathrm{LiV_{2}O_{4}}$ \cite{Kondo1997,Johnston2000},
and most recently Ca$_{2-x}$Sr$_{x}$RuO$_{4}$ with $x\sim 0.3$--0.5
\cite{Jin2001}.  There is currently no theoretical consensus on the mechanism for
formation of the heavy fermion mass in $\mathrm{LiV_{2}O_{4}}$ \cite{Fulde2004}.

An important measurement for establishing Fermi liquid behavior at low $T$
is nuclear magnetic resonance (NMR). For high magnetic purity samples of $%
\mathrm{LiV_2O_4}$, the $^7$Li nuclear spin-lattice relaxation rate $1/T_1$
is proportional to $T$ (the Korringa law for a Fermi liquid) from about 10 K down
to about 1.5 K \cite{Kondo1997,Mahajan1998,Fujiwara2004}.  In contrast to these
results, recent $^7$Li NMR measurements of several samples down to 30 mK by
Trinkl, Kaps et al.\ strongly conflict with a Fermi liquid
interpretation \cite{Trinkl2000,Kaps2001}. In particular, non-exponential
(stretched exponential) recovery of the nuclear magnetization, non-Korringa
behavior in $1/T_1$ versus $T$, a peak in $1/T_1$ at $\sim 0.6$ K, and a strong
field dependence of $1/T_1$ were found at low $T$\@. In view of the small number
of known $d$-electron HF compounds and the importance of $\mathrm{LiV_2O_4}$
within this small family, it is critical to determine if these results are
intrinsic to the pure material, and if not, what they are due to.

Here we present $^{7}$Li NMR measurements on two samples from 0.5 to 4.2 K
that were carried out to study the influence of magnetic defects on the low-$%
T$ HF properties of $\mathrm{LiV_{2}O_{4}}$. We confirmed Fermi liquid
behavior down to 0.5 K in a high magnetic purity sample. We find that a
small concentration (0.7 mol\%) of magnetic defects within the spinel
structure drastically changes the detected spin dynamics and leads to the
above behaviors described in Refs.\ \onlinecite{Trinkl2000} and %
\onlinecite{Kaps2001}, which therefore explain their results as arising from
a significant concentration of magnetic defects in their samples. On the
other hand, understanding the physics of magnetic defects in $\mathrm{%
LiV_{2}O_{4}}$ is interesting and important in its own
right and may further guide and constrain
theoretical models for the pure material. A crucial aspect of
\textrm{LiV}$_{2}$\textrm{O}$_{4}$ is the geometric frustration of V spins for
antiferromagnetic ordering in the spinel structure.  The geometric frustration is
likely directly related to the suppression of antiferromagnetic order in pure
\textrm{LiV}$_{2}$\textrm{O}$_{4}$ and the emergence of a heavy electron state
instead.  A large number of low lying spin excitations emerges and the system
becomes``almost unstable'', i.e.\ very susceptible with respect to crystal defects
or other perturbations that locally lift the frustration and cause a condensation
of the low lying states \cite{Millis2003}.

We develop a phenomenological description for the observed stretched
exponential $^7$Li nuclear relaxation in terms of a distribution of $1/T_1$
values  and explain the physical meaning of the parameters.  We further
present a model for defect nucleated dynamical magnetic order in an almost
unstable electronic system that can explain this distribution and that provides
important insights about the behaviors of magnetic defects in
$\mathrm{LiV_{2}O_{4}}$. The stretched exponential function also describes the
kinetics of diverse relaxation phenomena
\cite{Phillips1996}, so our model will likely have many applications to other
fields.

The two $\mathrm{LiV_2O_4}$ samples measured here were samples \#12-1 and
\#3-3-a2 studied previously in Ref.\ \onlinecite{Kondo1999}, where their
synthesis and characterization were described. The magnetic defect
concentration $n_{\mathrm{defect}}$ in the two samples was previously
estimated from magnetization measurements at low $T$ \cite{Kondo1999}.  Sample
\#12-1 shows a clear but weak intrinsic broad maximum in $\chi(T)$ at 16 K,
characteristic of high magnetic purity \cite{Kondo1997,Kondo1999}, with only a
tiny upturn in $\chi(T)$ below 4 K corresponding to $n_{\mathrm{defect}}$ = 0.01
mol\%. The second sample, \#3-3-a2, has $n_{\mathrm{defect}}$ = 0.7 mol\% which is
sufficiently large that the low-$T$ intrinsic broad maximum in $\chi(T)$ is
completely masked by the magnetic defect Curie-like term\cite{Kondo1999}.
The $^7$Li NMR measurements were performed with a Fourier transform (FT)
TecMag pulse spectrometer using $^4$He (1.5--4.2 K) and $^3$He (0.5--1.5 K)
cryostats. The $^7$Li NMR lineshape and the full width at half maximum
(FWHM) were obtained from the FT of half of the echo signal. The $^7$Li $1/T_1$
was determined by monitoring the recovery of the spin echo intensity
following a saturating pulse sequence of $\pi/2$ pulses. The typical $\pi/2$
pulse width was 2 $\mu s$. The measurements were carried out at a frequency
of 17.6 MHz (magnetic field $H$ = 10.6 kOe) so that a direct comparison of
our results with the corresponding $^7$Li NMR data at 17.3 MHz in Refs.\ %
\onlinecite{Trinkl2000} and \onlinecite{Kaps2001} could be made.

\begin{figure}[t]
\includegraphics[width=2.3in]{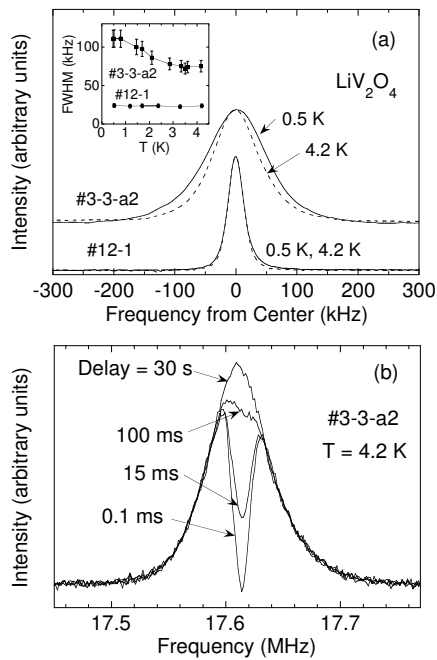}
\label{Fig1}\caption{(a) $^7$Li NMR lineshape at 0.5 and 4.2 K and full width 
at half maximum peak intensity versus temperature (FWHM, inset) for two samples of
${\rm LiV_2O_4}$.  (b) Hole-burning spectra at 4.2 K for sample \#3-3-a2.}
\end{figure}

The resonance line for the pure sample \#12-1 has a FWHM that is independent
of $T$ below 4.2 K whereas sample \#3-3-a2 has a much broader line [Fig.\
1(a)] that becomes increasingly broad with decreasing $T$ [inset, Fig.\
1(a)], indicating an increasing importance of magnetic inhomogeneity in the
latter sample with decreasing $T$\@. The Knight shift $K$ for both samples is
the same and independent of $T$ between 0.5 and 4.2 K, with $K$ =
0.141(15)\% in agreement with previous data \cite{Mahajan1998} above 1.5 K.

\begin{figure}[t]
\includegraphics[width=2.3in]{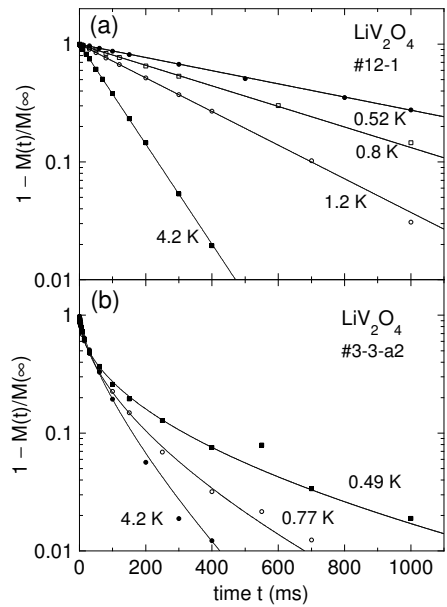} \label{Fig2}
\caption{Recovery of the $^7$Li nuclear magnetization $M$ vs time $t$ for $%
\mathrm{LiV_2O_4}$ samples \#12-1 (a) and \#3-3-a2 (b). The solid curves in
(a) and (b) are fits to the data at each $T$ by an exponential and a
stretched exponential, respectively.}
\end{figure}

Our main experimental results originate from measurements of the influence of
magnetic defects on the $^7$Li nuclear spin dynamics of $\mathrm{LiV_2O_4}$.
Figure 2(a) shows representative semilog plots of the time $t$ dependent recovery
of the
$^7$Li nuclear magnetization $M(t)$ after initial saturation for magnetically pure
sample \#12-1 at several $T$. The data at each $T$ lie on a straight line (shown)
with a well-defined $1/T_1$ determined from a fit of the data by $1 -
M(t)/M(\infty) = A$ exp$(-t/T_1)$, where the prefactor $A$ is typically 0.9 to
1.1. The resulting $1/T_1$ is plotted vs $T$ in Fig.\ 3. These data follow the
Korringa law for a Fermi liquid ($1/T_1 \propto T$) with a weighted fit giving
$(T_1T)^{-1} = 2.46(6) \mathrm{s^{-1}K^{-1}}$. Our results thus further confirm
Fermi liquid behavior for pure $\mathrm{LiV_2O_4}$ at low $T$.

The $M(t)$ for sample \#3-3-a2 with $n_{\rm defect}$ = 0.7 mol\% is shown
for representative temperatures in Fig.\ 2(b).  The recovery is drastically 
different from that of the pure sample, exhibiting strongly non-exponential
behavior.  Another important feature is that the shape of the recovery curve
changes with decreasing $T$, particularly strongly below 1 K\@.  Following Ref.\
\onlinecite{Kaps2001}, we fitted the data at each $T$ by the stretched
exponential function 
\begin{figure}[t]
\includegraphics[width=2.3in]{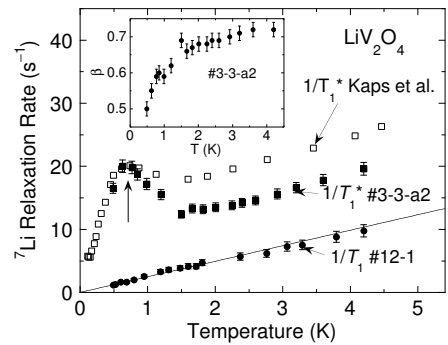}
\caption{$^7$Li nuclear spin-lattice relaxation rates in $\mathrm{LiV_2O_4}$
vs temperature $T$. The lowest data set of $1/T_1$ vs $T$ is for high purity
sample \#12-1; the fitted straight line is the Korringa law for a Fermi
liquid. The $1/T_1^*$ vs $T$ data sets are for our sample \#3-3-a2 (filled
squares) and from Kaps et al. (Ref. \protect\onlinecite{Kaps2001}) (open
squares). The vertical arrow at about 0.7 K indicates the approximate $T$ of the
maxima in $1/T_1^*$. The inset shows the exponent $\protect\beta$ in the stretched
exponential in Eq.\ (1) vs $T$.}
\end{figure}
\begin{equation}
1 - \frac{M(t)}{M(\infty)} = \mathrm{exp}[-(t/T_1^*)^\beta],  \label{Eq1}
\end{equation}
where $\beta$ is the stretching exponent with $0 < \beta \leq 1$ and $1/T_1^*$ 
is a characteristic relaxation rate.  This function is nonanalytic for
$t \rightarrow 0$.   We therefore employed a low-$t$ cutoff at 15 ms to the fit. 
The resulting $1/T_1^*(T)$ is plotted as filled squares in Fig.\ 3 and the fits
are shown by the solid curves in Fig.\ 2(b).  The variation of $\beta$ with $T$ is
shown in the inset of Fig.\ 3.  The $^7$Li NMR $1/T_1^*$ data obtained at
17.3 MHz by Kaps et al. \cite{Kaps2001} are plotted vs $T$ as open squares in
Fig.\ 3.  From the totality of the data in Figs.\ 1--3, we conclude that the data
of Kaps et al.\ are not intrinsic to pure ${\rm LiV_2O_4}$ but rather are
dominated by the influence of magnetic defects.

The non-exponential recovery of $M(t)$ in sample \#3-3-a2 suggests that there is a
distribution of $1/T_1$ values in this sample for different $^7$Li
nuclei.  To check this hypothesis, we performed relaxation measurements in 
different regions of the NMR spectrum.  These are shown in Fig.\ 1(b) as
``hole-burning'' experiments.  By using a long saturating $\pi/2$ pulse we can
irradiate only the central part of the spectrum.  The fact that the ``hole''
disappears gradually during relaxation without affecting the remaining part of
the line shape indicates the absence of spectral diffusion, which means that
$^7$Li nuclei with different Larmor frequencies have no thermal contact over our
time scale.  Another experiment was done by monitoring the recovery of the part of
the echo signal far from $t = 0$.  The recovery was found to be nearly
exponential and with a $1/T_1$ corresponding to the long time tail of the
stretched exponential in Fig.\ 2(b).  These two experiments together demonstrate
that there does exist a distribution of $1/T_1$ values for
the $^7$Li nuclei in the sample at each $T$ on our time scale and that
these nuclei or groups of nuclei relax independently.  The strong decrease of
$\beta$ with decreasing $T$ in the inset of Fig.\ 3 must therefore reflect a
significant change in the distribution of $1/T_1$ values with
$T$ as discussed next.

Our experiments thus demand that we model the stretched exponential relaxation in 
Eq.\ (1) as the sum over the sample of a probability distribution $P$ of $1/T_1$
for the various $^7$Li nuclei.   Accordingly we write the stretched exponential
function in Eq.\ (1) as
\begin{equation}
e^{-(t/T_{1}^{\ast })^{\beta }}=\int_{0}^{\infty }P(s,\beta
)e^{-st/T_{1}^{\ast }}ds\ ,  \label{Eq2}
\end{equation}%
where $s=T_{1}^{\ast }/T_{1}$ is the ratio of a particular relaxation rate $%
1/T_{1}$ within the sample to the fixed relaxation rate $1/T_{1}^{\ast }$
characteristic of $P(s,\beta )$, and of course $\int_{0}^{\infty }P(s,\beta
)ds=1$. For $\beta =1$, $P(s,1)$ is the Dirac delta function at $s=1$. For
general $\beta $, Eq.\ (2) shows that $P(s,\beta )$ is the inverse Laplace
transform of the stretched exponential. The evolution of $P(s)$ with $\beta $
for several values of $\beta $ is shown in Fig.\ 4.  With decreasing $\beta $,
$P(s)$ broadens and becomes highly asymmetric, and the peak in $P(s)$ becomes
finite and moves towards slower rates which is compensated by a long tail to
faster rates.  The value of $s$ at which $P(s)$ is maximum is plotted versus
$\beta $ in the inset of Fig.\ 4; this value decreases with decreasing $\beta $
and approaches zero exponentially for $\beta \lesssim 0.5$.  The
physical significance  of $1/T_{1}^{\ast }$ for $1/3\lesssim \beta <1$ is that
$1/T_{1}$ is about equally likely to be less than $1/T_{1}^{\ast }$ as it is to
be greater; it is neither the average of $1/T_{1}$ nor the inverse of the average
of $T_{1}$.  For $s\gg 1$, $P(s,\beta )\sim 1/s^{1+\beta }$, so the average
$s_{\mathrm{ave}} = (T_{1}^{\ast }/T_{1})_{\mathrm{ave}}$ is infinite.  Thus the
moments of the distribution depend on the cut-off at large relaxation rates, but
this cutoff is irrelevant for the physical interpretation of the long time
relaxation.  We see that the measured small values for $\beta$ at low $T$ in Fig.\
3 for sample \#3-3-a2 and that of Kaps et al.\ constitute strong evidence for a
broad distribution of $^7$Li $1/T_1$ values at these $T$.

\begin{figure}[t]
\includegraphics[width=2.3in]{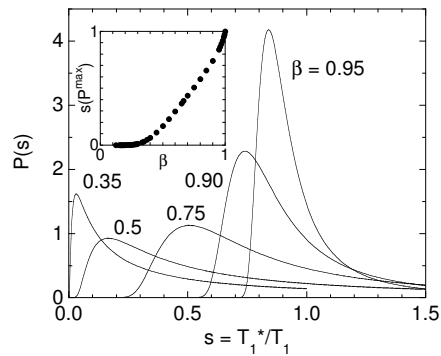}
\caption{Probability density $P(s)$ for several stretched exponential
exponents $\protect\beta$. The inset shows the $s$ values versus $\protect%
\beta$ at which $P(s)$ is maximum.}
\end{figure}

We now discuss possible physical origins of the $1/T_1$ probability distributions
in Fig.\ 4 and then propose a model that may be applicable to
$\mathrm{LiV_2O_4}$ containing magnetic defects.  We note at the outset that our
NMR measurements were carried out at $H=10.6$ kOe for which the magnetization of
the magnetic defects at $\lesssim 1$ K is significant \cite{Kondo1999}, a
situation for which very few calculations of either the average bulk or local
electronic spin fluctuations are available.  With this caveat, the contribution
to the $1/T_{1}$ of a nucleus located at site $\mathbf{r}$, by fluctuations of
electron spins $\mathbf{S}(\mathbf{r^\prime,}t)$ with correlation function
$q_{\mathbf{r^\prime,r^{\prime\prime}}}(t)=\langle S_{\alpha
}(\mathbf{r^\prime,}0)S_{\alpha }(\mathbf{r}^{\prime\prime }\mathbf{,}t)\rangle
$, is 
\begin{equation}
\frac{1}{T_{1} }\left( \mathbf{r}\right)\propto \int d^{3}r^{\prime
}d^{3}r^{\prime \prime }A_{\mathbf{r,r}^{\prime }}A_{\mathbf{r,r}^{\prime
\prime }}q_{\mathbf{r}^{\prime }\mathbf{,r}^{\prime \prime }}(\omega _{n}).
\label{T1}
\end{equation}%
Here $A_{\mathbf{r,r}^{\prime }}$ is the hyperfine interaction between
nuclear and electron spins at sites $\mathbf{r}$ and $\mathbf{r}^{\prime }$,
respectively, and $q_{\mathbf{r^\prime,r^{\prime\prime}}}(\omega _{n})$ is the
Fourier transform of the correlation function
$q_{\mathbf{r^\prime,r^{\prime\prime}}}(t)$ at the nuclear Larmor frequency
$\omega _{n}=\gamma _{n}H$.  For example, if one had a unique $1/T_1$, a unique
$A_{\mathbf{r,r}^{\prime }}$ with  $\mathbf{r^\prime}= \mathbf{r^{\prime\prime}}$,
and a correlation function $q_{\mathbf{r^\prime,r^{\prime}}}(t) \sim
e^{-\varepsilon t}$, one would have  $1/T_1 \propto \varepsilon/(\varepsilon^2 +
\omega_n^2)$, yielding a peak in $1/T_1$ as $\varepsilon$ decreases through
$\omega_n$ with decreasing $T$ as observed for $1/T_1^\ast$ at $\approx 0.7$ K in
Fig.\ 3.  This simplified example suggests that a significant fraction of the
magnetic defects drastically slows down below $\sim 1$ K.

A distribution of $1/T_{1}$ could result from a spatial variation in the
electron spin dynamics, i.e.\ $q_{\mathbf{r^\prime,r^{\prime\prime}}}(t)$, or
variations of the hyperfine interactions. In the latter case, a nonlocal
hyperfine interaction $A_{\mathbf{r,r}^{\prime }}\propto
\left\vert \mathbf{r-r}^{\prime }\right\vert ^{-3}$, caused by dipolar
and/or RKKY interactions, can lead to a broad distribution in $1/T_{1}$.
Depending on whether a given nuclear spin is close to or far away from a
local defect that dominates the spin response $q_{\mathbf{r,r}^{\prime }}(t)$,
very different $1/T_{1}$ values occur.  Geometric considerations lead to 
$P(s,\beta )\propto s^{-3/2}$, i.e.\ a fixed value $\beta =\frac{1}{2}$ for
the stretched exponential which is in direct conflict with our data that show a
strongly $T$-dependent $\beta$.  A physically more interesting case is 
when stretched exponential nuclear relaxation is due to dynamical heterogeneity
of the magnetic defect spin system, i.e. due to spatially varying
$q_{\mathbf{r^\prime,r^{\prime\prime}}}(t)$. For simplicity we consider the limit
of a purely local hyperfine interaction
$A_{\mathbf{r,r}^{\prime }}\propto \delta _{\mathbf{r-r}^{\prime }}$. In the
limit of strongly disordered spin systems, dynamical heterogeneity with anomalous
long time dynamics was found in numerical simulations above the spin glass
temperature \cite{Glotzer1998}. The averaged autocorrelation function was shown to
have the Ogielski form $q(t)=t^{-x}%
\mathrm{exp}[-(\varepsilon ^{\ast }t)^{\beta }]$, where
the energy scale
$\varepsilon ^{\ast }$ characterizes the averaged electron spin response.  As
$\varepsilon ^{\ast }$ becomes smaller than $\omega _{n}$ with decreasing $T$,
$1/T_{1}^\ast$ goes through a maximum, located at $T\sim 1\ \mathrm{K}$ in our
case, and $q\left( t\right) $ immediately yields a stretched exponential
relaxation for the nuclear spins.  However, these results
were obtained in the strong disorder limit in contrast to dilute magnetic defects
in $\mathrm{LiV_{2}O_{4}}$.

The sensitivity of the HF state with respect to perturbations may be
critical to understand the behavior of $\mathrm{LiV_{2}O_{4}}$ where
$n_{\mathrm{defect}}$ is rather small, suggesting the following alternative 
model for the dynamics.  If geometric frustration suppresses long
range order in pure \textrm{LiV}$_{2}$\textrm{O}$_{4}$, it becomes natural that
crystal defects can locally lift the frustration and cause a condensation of
dynamic magnetic order in a finite region of volume $\simeq \xi ^{3}$ around the
defect.  Due to the proximity to an ordered state and long-range electronic spin
coupling in metallic systems, we expect $\xi$ to be much larger than an
interatomic spacing, in contrast to insulating frustrated systems with only
nearest-neighbor interactions \cite{Villain1979}.  This might help to explain our
previous low-$T$ magnetization measurements which indicated that the magnetic
defects have large average spins $\sim $ 3/2 to 4 \cite{Kondo1999}.  Fluctuations
in the local tendency towards order lead to a probability $p\left(\xi \right)
\propto e^{-c\xi ^{3}}$ for such a droplet \cite{Millis02} and we analyze the
system using the ideas of Griffiths physics in disordered magnets
\cite{Griffiths69}.

The lowest excitation energies, $\varepsilon $, of a droplet depend
on its size $\xi$.  Depending on how $\varepsilon \left(\xi\right)$ varies with
$\xi$, different long-time dynamics emerges \cite{Vojta2005}.  If
$\varepsilon\propto \xi^{-\psi }$ the distribution function of the droplet
energies becomes $p\left( \varepsilon \right) \propto e^{-\left( \frac{\varepsilon
^{\ast }}{\varepsilon }\right) ^{3/\psi }}$. If $1/T_{1}\propto\varepsilon$, this
yields $P(s,\beta )\propto e^{-s^{-3/\psi }}$, leading for large times to a
stretched exponential relaxation with $\beta =\frac{1}{1+\psi /3}$. Such a
behavior occurs in magnets with Heisenberg symmetry \cite{Vojta2005} where one
finds $\beta =\frac{1}{2}$, if the spin dynamics is classical.  For lower $T$,
where quantum dynamics of the spins sets in, one finds $\beta
=\frac{2-z}{4-z}<\frac{1}{2}$, if the dynamical exponent relating length and time
scales obeys $z<2$.  For insulating antiferromagnets, one usually has $z = 1$,
giving $\beta = 1/3$.

Probably more appropriate to $\mathrm{LiV_2O_4}$ is the case of itinerant
antiferromagnets where one expects $z=2$, and an even more exotic situation
occurs. In this case,
$\varepsilon \propto e^{-b\xi ^{3}}$, i.e.\ large droplets become extremely slow
leading to quantum Griffiths behavior $P\left( s\right)
\propto s^{-\lambda }$ at long times with nonuniversal exponent $\lambda =1-c/b$.
Now, the nuclear spin relaxes according to a power law $1-M(t)/M(\infty )\propto
t^{-(1-\lambda )}$.  It becomes very hard to distinguish at large
times power law from stretched exponential behavior with small $\beta$ at low
$T$.  However, there are clear predictions of this scenario which include
(given $\lambda >0$) singular non-Fermi-liquid type specific heat
$C/T\propto T^{-\lambda }$, susceptibility $\chi \propto T^{-\lambda }$ and
similar results for the field dependence of the magnetization \cite{Vojta2005},
which can all be tested in future experiments. 

\begin{acknowledgments}
Discussions with T. Vojta are gratefully acknowledged.  Ames Laboratory is
operated for the U.S.\ Department of Energy by Iowa State University under
Contract No.\ W-7405-Eng-82. This work was supported by the Director for Energy
Research, Office of Basic Energy Sciences.
\end{acknowledgments}

\end{document}